# Heavy-tailed open quantum systems reveal long-lived and ultrasensitive coherence

Sunkyu Yu[1†], Xianji Piao[2§], and Namkyoo Park[3*]

[1]Intelligent Wave Systems Laboratory, Department of Electrical and Computer Engineering, Seoul National University, Seoul 08826, Korea

[2]Wave Engineering Laboratory, School of Electrical and Computer Engineering, University of Seoul, Seoul 02504, Korea

[3]Photonic Systems Laboratory, Department of Electrical and Computer Engineering, Seoul National University, Seoul 08826, Korea Seoul 08826, Korea

E-mail address for correspondence: sunkyu.yu@snu.ac.kr; piao@uos.ac.kr; nkpark@snu.ac.kr

**Abstract**

Understanding random open quantum systems is critical for characterizing the performance of large-scale quantum devices and exploring macroscopic quantum phenomena. Various features in these systems, including spectral distributions, gap scaling, and decoherence, have been examined by modelling randomness under the central limit theorem. Here, we investigate random open quantum systems beyond the central limit theorem, focusing on heavy-tailed system-environment interactions. By extending the Ginibre unitary ensemble, we model system-environment interactions to exhibit a continuous transition from light-tailed to heavy-tailed distributions. This generalized configuration reveals unique properties—gapless spectra, Pareto principle governing dissipation, orthogonalization, and quasi-degeneracies—all linked to the violation of the central limit theorem.

The synergy of these features challenges the common belief—the tradeoff between stability and sensitivity—through the emergence of long-lived and ultrasensitive quantum coherences that exhibit an enhancement of two orders of magnitude compared to predictions under the central limit theorem. The result, which is based on heavy-tailedness of open quantum systems, provides highly desirable platforms for quantum sensing applications.

# Introduction

The central limit theorem (CLT)—which states the convergence of a broad range of random distributions to the normal distribution—enables universal analysis of diverse random processes within a unified framework[1]. On the other hand, the importance of exploring random distributions that violate the CLT has been demonstrated across various fields, from macroscopic to microscopic systems. Notable examples include heavy-tailed distributions[2], which play crucial roles in the Gutenberg-Richter law in seismology[3], hub-dominant signal transport over scale-free networks[4], efficient training of artificial neural networks[5], optimization of photonic circuits[6], Lévy flights of light[7], and energy spectra in quantum turbulence[8].

Along with efforts to exploit high-dimensional[9] and complex[10] quantum systems, the CLT has been applied to uncover universal features of random open quantum systems[11-16]. In these approaches, the system Hamiltonians and their interactions with environments are modelled using a normal-distribution family of random matrices, such as the Gaussian unitary ensemble (GUE)[17], the Ginibre unitary ensemble (GinUE)[18], and the Wishart-Laguerre unitary ensemble (WLUE)[19]. This CLT-based random matrix theory has uncovered various universal features in open quantum systems—bounded spectral distributions[12], gap vanishing scaling with system dimensionality[11], separated relaxation time scales under strong dissipation[13], symmetry-restricted spectra[20], locality-induced spectral changes[14], and universal decoherence rates[16].

Given the successes of exploring the CLT breaking in modelling physical phenomena[3,4,7,8] and achieving advanced functionalities[5,6], it is imperative to generalize random open quantum systems beyond the CLT. Very recently, a study implies the potential of this perspective by demonstrating the alteration of relaxation via power-law deformations of Gaussian ensemble operators[21]. A similar approach also reveals the impact of rare events in decoherence processes by exploring an ensemble of two-level fluctuators[22]. However, a comprehensive exploration of universal tail-dependent behaviours, and especially, their relation to CLT violation remain a challenge.

Here, we examine Markovian open quantum systems characterized by heavy-tailed distributions of random system-environment interactions. We employ the Student's *t* distribution to model these interactions, which enables a continuous transition between light-tail and heavy-tail regimes. By solving the Lindblad quantum master equation, we identify unique features of our heavy-tailed systems, including tail-dependent gap narrowing, the Pareto principle for quantum dissipation, the recovery of state orthogonality, and quasi-degeneracies. These spectral features, reflecting low-dissipation bulk states and high-dissipation outliers, lead to counterintuitive quantum dynamics—long-lived quantum coherence with ultrasensitive responses to random perturbations. These results provide an insight into the engineering of system-environment interactions[23] for quantum sensing and memory applications.

## Results

**Heavy-tailed open quantum systems**

To investigate open quantum systems evolving under the Markovian approximation, we start with the Gorini-Kossakowski-Sudarshan-Lindblad (GKSL) equation in its first standard form[24,25]:

$$\frac{d\rho}{dt} = -\frac{i}{\hbar}[H,\rho] + \left(\frac{\alpha}{\hbar}\right)^2 \sum_{k,l} \gamma_{kl} \left( S_l \rho S_k^\dagger - \frac{1}{2}\{S_k^\dagger S_l, \rho\} \right), \qquad (1)$$

where $\rho$ is the system density operator; $\alpha$ is the overall strength of system-environment interactions; $H$ is the system Hamiltonian including the Lamb shift; $\{S_k\}$ is the set of interaction operators acting on the system Hilbert space; and $\gamma_{kl}$ are operator-dependent interaction coefficients, constituting the positive semidefinite (PSD) Kossakowski matrix $K$ via $[K]_{kl} = \gamma_{kl}$. Equation (1) yields a completely positive and trace-preserving (CPTP) linear map for the evolution of an $N$-dimensional system, the interactions of which with the environment are fully specified by the $N^2 - 1$ traceless basis matrices $\{S_k\}$ together with the $(N^2 - 1) \times (N^2 - 1)$ matrix $K$. Therefore, open quantum systems subject to random system-environment interactions can be characterized by an ensemble of random realizations of $\{S_k\}$ and $K$.

Due to its PSD nature, $K$ can be expressed as a Gram matrix[12,26], $K = NXX^\dagger/\text{Tr}[XX^\dagger]$, which is normalized with $\text{Tr}[K] = N$ for a fair comparison between different realizations. In view of the CLT, a natural choice for $X$ is a complex square Ginibre matrix—that is, the matrix sampled from the GinUE[18]—with independent complex Gaussian entries (see Methods). Notably, the Gram-matrix configuration of the GinUE $X$ leads to $K$ sampled from the WLUE[19,26]. Due to the unitary invariance of the obtained $K$, the statistical nature of WLUE-sampled open quantum systems is preserved against basis transformations of $\{S_k\}$, thereby allowing investigation of universal features under the CLT[12].

To explore regimes beyond the CLT, we apply heavy-tailed distributions to $X$ instead of the light-tailed Gaussian distribution of the GinUE. In analysing the effect of tail thicknesses, we preserve the other characteristics of the GinUE $X$; the real and imaginary parts of its entries remain independent, identically distributed, and symmetric. Especially, to examine the continuous transition between light-tailed and heavy-tailed distributions, we employ the Student's $t$ distribution[27] for $X$, by using the probability density function (PDF):

$$p(x) = \sqrt{\frac{2}{\nu\pi}} \frac{\Gamma\left(\frac{\nu+1}{2}\right)}{\Gamma\left(\frac{\nu}{2}\right)} \left(1 + \frac{2}{\nu}x^2\right)^{-\left(\frac{\nu+1}{2}\right)}, \tag{2}$$

where $x = \text{Re}[X_{ij}]$ or $\text{Im}[X_{ij}]$ denotes the real or imaginary entries of $X$, $\Gamma$ is the gamma function, and $v > 0$ is the key parameter determining the tail thickness. Because only moments of order $r < v$ are finite[27], we distinguish the statistical phases of $X$ and the resulting Kossakowski matrix $K = NXX^\dagger/\text{Tr}[XX^\dagger]$ by controlling $v$ (Fig. 1a): the "Extremely Heavy Tail (EHT)" regime with undefined or infinite variance ($r = 2$) for $0 < v \leq 2$, the "Heavy Tail (HT)" regime with finite variance and infinite kurtosis ($r = 4$) for $2 < v \leq 4$, the "Moderately Heavy Tail (MHT)" regime with finite variance and kurtosis but heavier than a Gaussian distribution for $4 < v \lesssim 30$, and convergence to the light-tailed (LT) Ginibre matrix $X$ for $v \gtrsim 30$ with normally distributed real and imaginary entries[28]. Figures 1b and 1c illustrate the transition from the LT to the EHT regimes by controlling $v$, comparing theoretical (dashed lines) and numerical (dots) estimations of the PDF for $\text{Re}[X_{ij}]$. Notably, CLT violation arises only in the EHT regime, despite the presence of heavy tails also in the HT and MHT regimes.

In contrast to the WLUE realizations[12], constructing $K$ with the $t$ distribution no longer guarantees statistical invariance against the transformation of the basis $\{S_k\}$. However, owing to regularly varying Student's-$t$ tails and the principle of the single big jump[29], the symmetric power-law tails $p(|x|) \sim |x|^{-(v+1)}$ are preserved under arbitrary basis transformations (see Methods). Consequently, our modelling of open quantum systems allows the examination of universal tail behaviours independent from the chosen basis $\{S_k\}$. In the following analysis, we employ the standard SU($N$) generator basis[30] normalized to unity Hilbert-Schmidt norm, $\text{Tr}[S_k^\dagger S_l] = \delta_{kl}$.

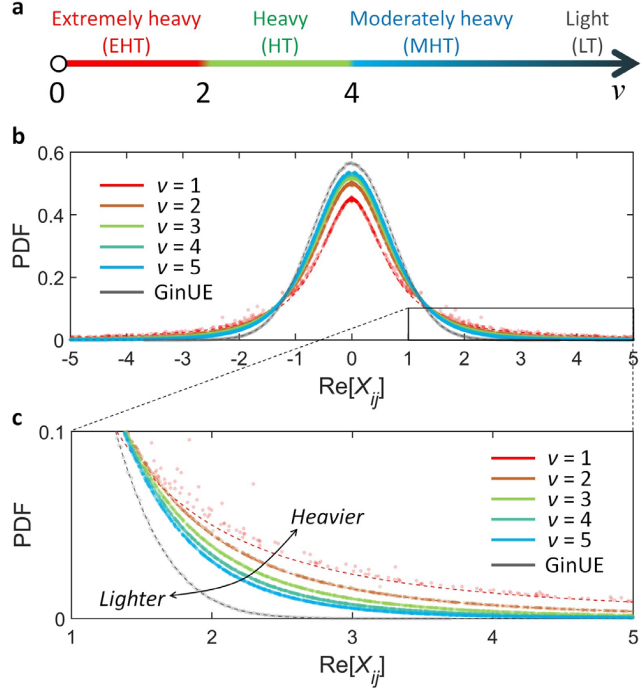

**Fig. 1. Heavy-tailed system-environment interactions. a,** $v$-dependent phases of the Student's $t$ distributions determined by their tail thicknesses. **b,c,** The entire PDFs of $\text{Re}[X_{ij}]$ for the $t$ distributions with different $v$'s and the GinUE (**b**) and their enlarged views near the tails (**c**). Dots and dashed lines in (**b,c**) denote numerical and theoretical estimations of the PDFs. The PDFs for 64 realizations are overlaid in (**b,c**). $N = 50$ for all cases.

**Tail dependence**

Using $K$ generated from the $t$ distribution, we investigate the tail dependence of random open quantum systems, first focusing on their eigenspectra and eigenstates. In the Liouville-space representation[31], Eq. (1) for an $N$-dimensional system becomes $d|\rho\rangle/dt = L|\rho\rangle$, where $|\rho\rangle \in \mathcal{L}_N$ denotes the $N^2 \times 1$ column-vectorized superket of the density operator $\rho$ in the Liouville space $\mathcal{L}_N$, and the $N^2 \times N^2$ matrix $L$ is the non-Hermitian Liouville superoperator (see Methods for obtaining $L$). For this non-Hermitian equation, the right-eigenvalue problem, $L|\rho_m^R\rangle = \lambda_m|\rho_m^R\rangle$, yields complex eigenvalues $\lambda_m$ for right eigenstates $|\rho_m^R\rangle$. We order these $N^2$ eigenvalues in descending

order of their real parts, as $\text{Re}[\lambda_m] \geq \text{Re}[\lambda_{m+1}]$. Because there is always a maximally mixed invariant state[24] $\rho_1 = I_N / N$ with $\lambda_1 = 0$, where $I_N$ is the $N$-dimensional identity matrix, and $\text{Re}[\lambda_m] \leq 0$ for all $m$, this ordering defines the spectral gap[12,32], $\Delta\lambda \triangleq \text{Re}[\lambda_1] - \text{Re}[\lambda_2] = -\text{Re}[\lambda_2]$. Notably, $\Delta\lambda$ determines the slowest relaxation dynamics toward the maximally mixed state[32].

In addition to eigenspectral properties, we examine eigenstates in terms of their nonorthogonality. With the left-eigenvalue problem, $\langle\rho_m^L|L = \lambda_m\langle\rho_m^L|$, we set $|\rho_m^R\rangle$ and $|\rho_m^L\rangle$ to form a biorthonormal basis of $L$, satisfying $\langle\rho_m^L|\rho_n^R\rangle = \delta_{mn}$. The Petermann factor, which quantifies the nonorthogonality of the $m$th biorthonormal eigenstates, is defined as[33,34]

$$K_m = \langle\rho_m^R|\rho_m^R\rangle\langle\rho_m^L|\rho_m^L\rangle = \text{Tr}\left[\rho_m^{R\dagger}\rho_m^R\right]\text{Tr}\left[\rho_m^{L\dagger}\rho_m^L\right], \qquad (3)$$

where $\rho_m^R$ and $\rho_m^L$ denote the $N \times N$ matrix representations of the eigen-superkets $|\rho_m^R\rangle$ and $|\rho_m^L\rangle$, respectively. The Petermann factor $K_m$ is a real number with $K_m \geq 1$, where $K_m = 1$ for all $m$ corresponds to a Hermitian $L$ with an orthonormal basis.

In examining $\Delta\lambda$ and $K_m$, we first consider purely dissipative systems by setting $H = O$ to isolate the effect of system-environment interactions. Figure 2 shows the complex-plane distributions of $\lambda$ for statistically distinct ensembles of $X$, where $\text{Re}[\lambda]$ and $\text{Im}[\lambda]$ govern the dissipative and oscillatory behaviours of quantum dynamics, respectively. For an ensemble from each $v$-dependent $t$ distribution, we examine 64 realizations generated independently. For comparison, we reproduce the result using GinUE $X$ (Fig. 2a), which exhibits the "lemon-like" $\lambda$-distribution[12] with $\lambda_1 = 0$ and $\Delta\lambda \approx 1 - 2/N$. Within this distribution, the relatively long-lived oscillatory states with high quality factor, $Q \triangleq |\text{Im}[\lambda]/(2\text{Re}[\lambda])|$, appear at the upper and lower extremes of the spectrum, marking the upper $Q$-factor bound (blue dashed lines in Fig. 2a-f). In all figures, the Petermann factor $K_m$ of each state is represented by the colour of its marker.

Figures 2b-2f show the transition of spectral distributions as the tail parameter $v$ varies across the HT and EHT regimes (Supplementary Note S1 for the MHT regime and for full-range plots at $v = 2$ and 3). Although the familiar lemon-shaped distribution remains stable until $v > 4$, the distribution becomes distorted in the HT regime with infinite kurtosis ($2 < v \leq 4$), which is most noticeably on the left side of the distribution: $\text{Re}[\lambda] < -1$ (Fig. 2b,c). Despite this change, key features of the GinUE realizations—the presence of a spectral gap, the preservation of the GinUE $Q$-factor bound, and the overall level of $K_m$ values—remain essentially unchanged in the HT regime, where the CLT still holds. It is also worth mentioning that the distribution in Fig. 2c captures the intrinsic nature of the $t$-distribution as a scale mixture of Gaussians[1] (see Methods).

We note that dramatic changes particularly on the right side of the distribution emerge as the system enters the EHT regime with infinite variance, where the CLT breaks down. First, in contrast to the GinUE reference (Fig. 2a) or the HT regime (Fig. 2b,c), the EHT regime with broken CLT exhibits a pronounced narrowing of the spectral gap (Fig. 2d)—which ultimately vanishes ($\Delta\lambda \to 0$) near $v = 1$ (Fig. 2e)—and the emergence of highly dissipative states ($\text{Re}[\lambda] \ll -1$ in Fig. 2f). Second, the overall level of $K_m$ appears to decrease substantially at $v = 1$ and 2 regardless of the values of $\lambda$, which indicates the recovery of orthogonality in the open system. In addition, the GinUE $Q$-factor bound (blue dashed lines) is violated with more frequent occurrences of high-$Q$ eigenstates, as analogous to the frequent emergence of hub nodes in heavy-tailed networks[35].

To generalize the results observed in purely dissipative systems, we extend our analysis to random system Hamiltonians $H$ sampled from the GUE (see Methods and Supplementary Note S2). Despite some discrepancies, such as the substantial broadening of $\text{Im}[\lambda]$ for smaller $\text{Re}[\lambda]$, the results follow trends similar to Fig. 2: gap narrowing, high-dissipation states, reduced Petermann factors, and *Q*-bound violation with heavier tails, confirming the statistical universality of these heavy-tail properties.

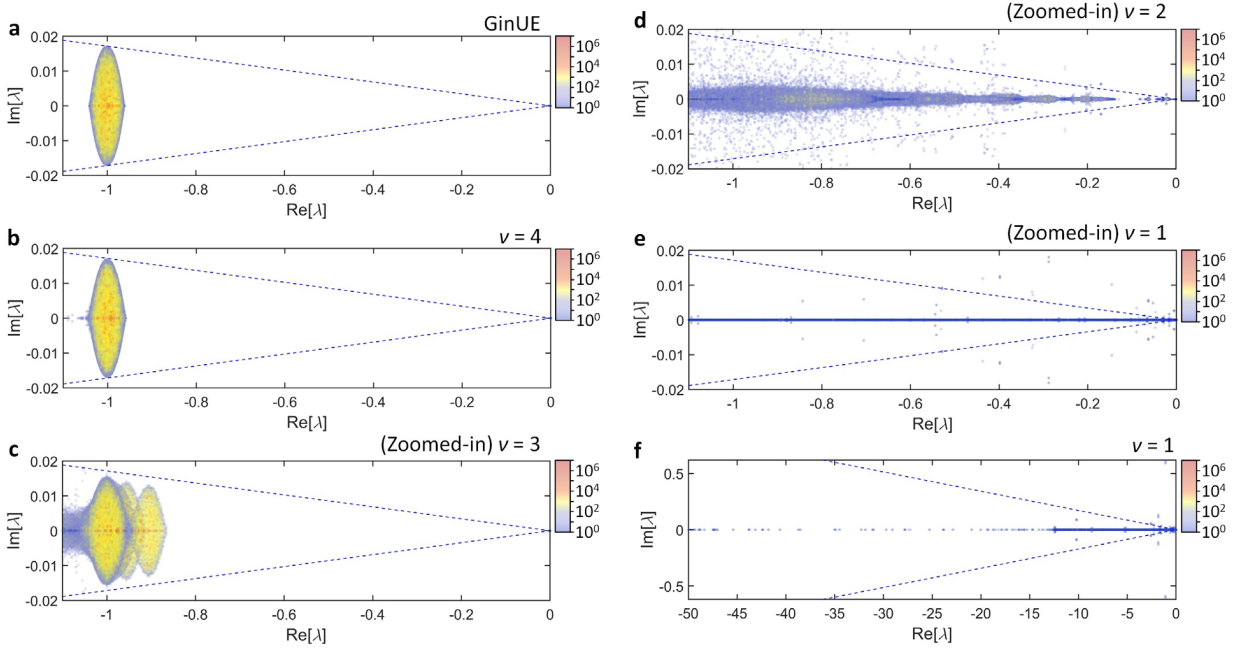

**Fig. 2. Spectral and state properties of purely dissipative systems.** Eigenvalue distributions for GinUE (**a**), and for the ensembles generated using the Student's *t* distributions with $v = 4$ (**b**), $v = 3$ (**c**), $v = 2$ (**d**), and $v = 1$ (**e,f**). Panels (**c-e**) employ the same plot range as panels (**a,b**), focusing on values near $\lambda = 0$; the full range plots for (**c,d**) appear in Supplementary Note S1. The colours of markers represent the $K_m$ values. For each ensemble, 64 realizations of $N = 50$ systems are examined, resulting in 160,000 eigenvalues overlaid as points on the complex plane. Blue dashed lines highlight the GinUE upper bound of eigenstate *Q* factor by extending from the origin. We set $\alpha = 1$ and $\hbar = 1$.

**Relaxation and sensitivity**

The observed spectral and state properties determine the quantum dynamics governed by Eq. (1). While the spectral gap $\Delta\lambda$ dictates the slowest relaxation of quantum features, such as coherences, the overall distribution of Re[$\lambda$] governs the averaged relaxation dynamics. Therefore, we analyse the tail dependence of $\Delta\lambda$ values and Re[$\lambda$] distributions in purely dissipative systems (see

Supplementary Note S3 for the extended distribution data and Supplementary Note S4 for another unique feature—$Q$-bound violation). Consistent with Fig. 2, $\Delta\lambda$ undergo a sharp transition across CLT violation threshold $v = 2$ (Fig. 3a). Especially, at $v = 1$, we observe a two-order-of-magnitude reduction in $\Delta\lambda$, which will substantially extend the upper bound of the relaxation time. Beyond this slowest relaxation dynamics, the complementary cumulative distribution functions (CCDFs) of Re[$\lambda$] (Fig. 3b) predict a time-domain extension also in the averaged relaxation behaviour— over 60% ($v = 2$) to 80% ($v = 1$) of states in the EHT regime have Re[$\lambda$] $\geq -1$, in sharp contrast to only about 50% of states in the HT and MTH regimes, and in GinUE realizations. Consequently, the spectral properties of open quantum systems violating the CLT indicate the emergence of long-lived quantum states in both their slowest and averaged relaxation dynamics. We note that these long-lived states originate from the heavy-tailed nature of the Kossakowski matrix $K$, which leads to high-dissipation outliers (Re[$\lambda$] < $-1$) and a low-dissipation bulk (Re[$\lambda$] $\geq -1$), especially in accordance with the 20:80 Pareto principle at $v = 1$.

In addition to relaxation dynamics, we can also predict the perturbative sensitivity of dynamics, which distinguishes the appropriate applications of a system between quantum computation and sensing. Remarkably, the Petermann factor $K_m$ of a quantum state is linked to its sensitivity to system perturbations[34]. Upon introducing a perturbation to the system Hamiltonian, $H = H_0 + V$, the eigenvalue shifts up to second order become (Methods)

$$\delta\lambda_m = \left\langle \rho_m^{\mathrm{L}} \middle| \Delta L \middle| \rho_m^{\mathrm{R}} \right\rangle + \sum_{n \neq m} \frac{\left\langle \rho_m^{\mathrm{L}} \middle| \Delta L \middle| \rho_n^{\mathrm{R}} \right\rangle \left\langle \rho_n^{\mathrm{L}} \middle| \Delta L \middle| \rho_m^{\mathrm{R}} \right\rangle}{\lambda_m - \lambda_n}, \quad (4)$$

where $\Delta L = I_N \otimes V - V^{\mathrm{T}} \otimes I_N$ denotes the resulting perturbation of the superoperator $L$. Because the first term is bounded by $|\langle \rho_m^{\mathrm{L}} | \Delta L | \rho_m^{\mathrm{L}} \rangle| \leq K_m^{1/2} \|\Delta L\|$ according to the Cauchy–Schwarz inequality, where $\|\Delta L\|$ denotes the spectral norm of $\Delta L$, $K_m$ characterizes the achievable first-order sensitivity. However, when considering the second-order perturbation, we note that the distances between eigenvalues, $\lambda_m - \lambda_n$, are also critical for evaluating the overall sensitivity. Therefore, to analyse the perturbative sensitivity of quantum dynamics in random open quantum systems, we examine the Petermann factor and the spacing between eigenvalues for varying $v$. In accordance with Eq. (4), we estimate the eigenvalue spacing via nearest-neighbour eigenvalue difference $\delta\lambda_{\text{NN}} = \min_{n \neq m}(|\lambda_m - \lambda_n|)$ evaluated over all $m$.

Although the averaged Petermann factors $\langle K_m \rangle$ experience a similar transition with $\Delta\lambda$ across CLT violation (Fig. 3c versus 3a), the mechanism governing sensitivity is far more intricate. We note that while the decreasing Petermann factors—accompanying the recovery of state orthogonality—suggest reduced first-order perturbative sensitivity, Fig. 3d reveals that heavier-tailed open quantum systems exhibit a one- to two-order-of-magnitude increase in near-degenerate eigenvalues ($\delta\lambda_{\text{NN}} \approx 10^{-5}$). This increase stems from the transition of the $\delta\lambda_{\text{NN}}$ distribution from Gaussian (GinUE) to a power-law ($v = 1$), in excellent agreement with the Pareto-principle separation of high- and low-dissipation states and the resulting highly concentrated low-dissipation bulk at $v = 1$. Because near-degeneracies enhance higher-order perturbative sensitivity according to Eq. (4), the contrasting behaviours of $K_m$ and $\delta\lambda_{\text{NN}}$ lead to a competition between first- and higher-order sensitivities under CLT violation. In Supplementary Note S5, we demonstrate the universality of the observed features with purely dissipative systems, by examining open systems with GUE Hamiltonians for relaxation and sensitivity.

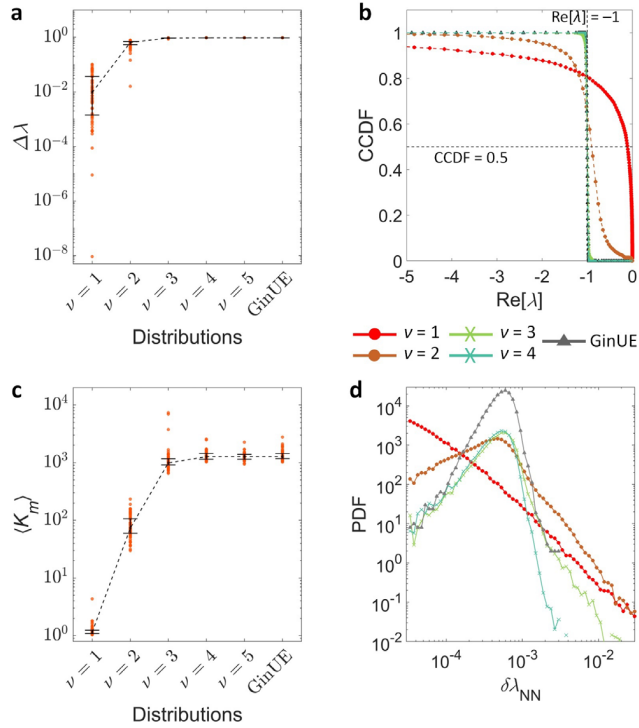

**Fig. 3. Tail-dependent relaxation and sensitivity parameters of purely dissipative systems. a,b,** Relaxation parameters: spectral gaps $\Delta\lambda$ (**a**) and CCDFs of Re[$\lambda$] (**b**). **c,d,** Sensitivity parameters: the averaged Petermann factors $\langle K_m \rangle$ (**c**) and PDFs of nearest-neighbour eigenvalue spacings $\delta\lambda_{NN}$ (**d**). In (**a-d**), $\nu$-dependent heavy-tailed systems are compared with GinUE realizations. In (**a,c**), each marker represents one ensemble realization, and error bars show the median (black dashed lines), and the first and third quartiles. All the other parameters are the same as those in Fig. 2.

**Coherence dynamics**

To verify the predictions of Figs. 2 and 3—the emergence of long-lived quantum states and the competition between nonorthogonality and quasi-degeneracy in sensitivity—we investigate quantum dynamics in random systems: the GUE Hamiltonians $H_0$ and the initial states $\rho(t = 0)$ sampled from Haar random pure states (see Methods). The dynamics are computed using system propagators obtained from the biorthonormal eigenstates and eigenvalues of $L$. To probe the relaxation properties of quantum states, we calculate their quantum coherence defined by the relative entropy $C_E(\rho) = S(\Delta_\rho) - S(\rho)$, where $S(\rho) = -\text{Tr}[\rho \log \rho]$ is the von Neumann entropy and $\Delta_\rho$ is obtained by discarding all off-diagonal entries of $\rho$ (Supplementary Notes S6 for the time evolutions of $C_E$ and $S$).

From the obtained temporal dynamics of $C_E$, we estimate the coherence time $T_2$ (see Methods) as a relaxation metric. To examine sensitivity, we introduce a Gaussian random perturbation, $H = H_0 + V$, where $V$ is sampled from the GUE with $\langle \text{Tr}[V] \rangle = \langle \text{Tr}[H_0] \rangle / 10$. Sensitivity to the perturbation is then estimated by two metrics: the alteration in coherence time $\Delta T_2$ and the time-averaged coherence perturbation $\Delta C_E$, which is defined by

$$\Delta C_E = \frac{1}{2T_2} \int_0^{2T_2} \left( \ln \frac{C_E^V(t)}{C_E^0(t)} \right)^2 dt, \tag{5}$$

where $C_E^0(t)$ and $C_E^V(t)$ denote the original and perturbed time evolutions of coherence, respectively.

Figure 4 illustrates the tail dependence of these metrics for quantum dynamics. We note that the EHT regime with CLT violation yields surprisingly long-lived coherence with $T_2$ increased by one to three orders of magnitude (Fig. 4a), as expected from the substantial gap narrowing and the increase of a low-dissipation bulk in this regime (Fig. 3a,b). Remarkably, this long-lived coherence is highly sensitive to perturbations, as rigorously demonstrated by both $\Delta T_2$ (Fig. 4b) and $\Delta C_E$ (Fig. 4c). These results indicate that among two mechanisms—state nonorthogonality and quasi-degeneracy—of perturbative sensitivity, the quasi-degeneracy, and therefore, higher-order perturbations dominate the system response, which originate from the concentration of lower-dissipation states ($\text{Re}[\lambda] \geq -1$). Consequently, open quantum systems arising from extremely heavy tails in system-environment interactions possess quantum states that are simultaneously long-lived and highly sensitive, with metrics extended by roughly two orders of magnitude.

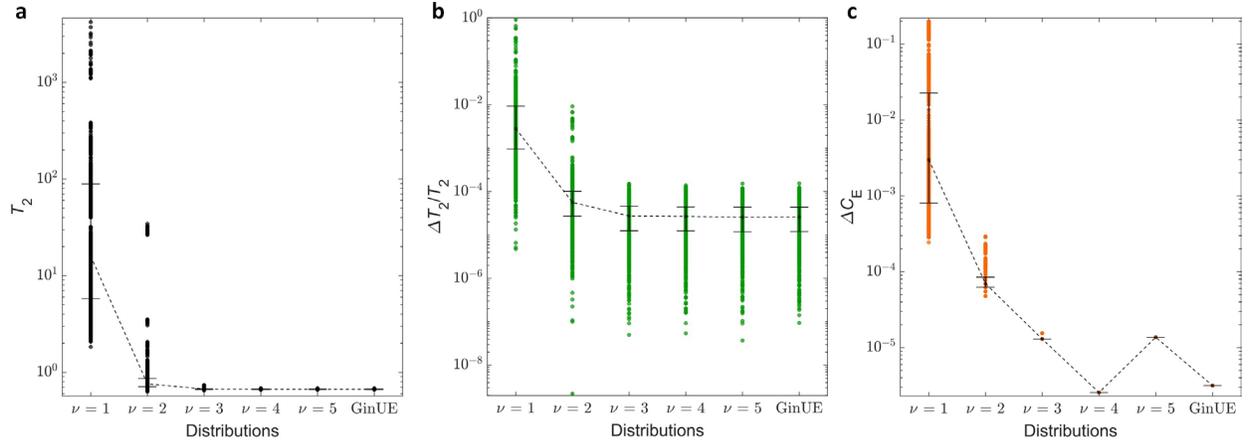

**Fig. 4. Coherence time and sensitivity metrics in random open systems. a,** Coherence time $T_2$. **b,c,** Sensitivity metrics: coherence time perturbation $\Delta T_2$ (**b**) and time-averaged coherence perturbation $\Delta C_E$ (**c**). Each marker represents one ensemble realization characterized by a tail-dependent $K$, a GUE Hamiltonian $H_0$ (**a**) or the pair of $H_0$ and $H_0 + V$ (**b,c**), and a Haar random pure initial state. Error bars show the median (black dashed lines), and the first and third quartiles for 1,280 realizations obtained from 64 open quantum systems and 20 initial states. All the other parameters are the same as those in Fig. 2.

## Discussion

The superior coherence times observed in our heavy-tailed open quantum systems are fundamentally distinct from those of other long-lived quantum states, which rely on suppressed system-environment interactions[36], symmetrized[37,38] or correlated[39,40] interactions, or frequent measurements[41]. In our study, we preserve the overall strength of system-environment interactions by normalizing the Kossakowski matrix as $\text{Tr}[K] = N$, thereby separating the impact of statistical distributions from the interaction strength[36]. Furthermore, the proposed heavy-tailed systems possess neither symmetries[37,38], correlations[39,40], nor dynamical interactions[38,41], yet attaining enhanced coherence under static random interactions, provided that the interactions follow heavy-tailed distributions with CLT violation. Therefore, our findings reveal a novel physical origin of long-lived quantum states, particularly relevant for complex and high-dimensional quantum systems.

We note that the counterintuitive achievement of highly sensitive behaviours with stable coherence is desirable for quantum sensing[42]. Our findings can be understood as the extension of the exceptional-point-enhanced sensing[34,43] to open quantum systems, by exploiting quasi-degenerate states instead of the exceptional point for enhanced sensitivity, and gap annihilation for enhanced coherence. Because our model lacks any symmetry, the observed result also motivates the generalization of parity-time-symmetric sensing to non-Hermitian disordered platforms in both classical and quantum regimes.

In conclusion, we investigated how heavy-tailed distributions in system-environment interactions shape the Lindblad dynamics of open quantum systems. Compared to random open quantum systems governed by the CLT, the CLT violation with extremely heavy tails derives various unique features: spectral gap narrowing, Pareto-type high-dissipation outliers and low-dissipation bulk, state orthogonalization, and reduced eigenvalue spacing. We observed that, as the spectrum approaches the gapless regime, the systems exhibit orders-of-magnitude enhancement in both coherence times and perturbative sensitivity, which are desired properties for quantum sensing. The implementation of quantum many-body systems toward CLT violation and the extension of our approach from statistical distributions to network properties would be future research topics.

## Methods

**Ginibre unitary ensemble.** As a comparison group to our heavy-tailed open quantum systems, we prepare the Wishart Kossakowski matrix—that is, the matrix sampled from the WLUE—which was examined for obtaining universal spectral properties of random open quantum systems under the CLT[12]. The Wishart matrix is obtained as $K = NXX^\dagger/\text{Tr}[XX^\dagger]$ using the complex matrix $X$ sampled from the GinUE, where the real and imaginary entries of $X$, $x = \text{Re}[X_{ij}]$ or $\text{Im}[X_{ij}]$, are independent Gaussian random variables $\mathcal{N}(0, 1/2)$ with the following PDF:

$$p(x) = \frac{1}{\sqrt{\pi}} e^{-x^2}. \tag{6}$$

**Universality in tail behaviours.** In constructing the Kossakowski Gram matrix $K = NXX^\dagger/\text{Tr}[XX^\dagger]$ for Eq. (1), we define each entry $X_{ij}$ of $X$ to be a complex random variable of which the real and imaginary parts independently follow the Student's $t$ distribution given by Eq. (2). According to the representation of the $t$ distribution as a scale mixture of Gaussian distributions[1], $X_{ij}$ can be written as

$$\text{Re}[X_{ij}] = \frac{1}{\sqrt{2}} \frac{G_{ij}^{(R)}}{\sqrt{W_{ij}^{(R)}/\nu}}, \quad \text{Im}[X_{ij}] = \frac{1}{\sqrt{2}} \frac{G_{ij}^{(I)}}{\sqrt{W_{ij}^{(I)}/\nu}}, \tag{7}$$

where $G_{ij}^{(R)}$ and $G_{ij}^{(I)}$ are independent standard Gaussian random variables, as $G_{ij}^{(R,I)} \sim \mathcal{N}(0, 1)$, and $W_{ij}^{(R)}$ and $W_{ij}^{(I)}$ are independent chi-squared random variables with $\nu$ degrees of freedom, as $W_{ij}^{(R,I)} \sim \chi_\nu^2$. All these random variables are mutually independent.

Because the basis transformation of $\{S_k\}$ in Eq. (1) leads to the unitary transformation of the Kossakowski matrix[24,25], $K \to K^{(U)} = UKU^\dagger = NUX(UX)^\dagger/\text{Tr}[XX^\dagger]$, the universality of our study is governed by the invariant features under the transformation $X \to X^{(U)} = UX$. From Eq. (7), we obtain the unitarily transformed entries, as

$$\mathrm{Re}\left[X_{ij}^{(U)}\right] = \frac{1}{\sqrt{2}} \sum_k \left( \frac{\mathrm{Re}[U_{ik}] G_{kj}^{(R)}}{\sqrt{W_{kj}^{(R)}/v}} - \frac{\mathrm{Im}[U_{ik}] G_{kj}^{(I)}}{\sqrt{W_{kj}^{(I)}/v}} \right),$$

$$\mathrm{Im}\left[X_{ij}^{(U)}\right] = \frac{1}{\sqrt{2}} \sum_k \left( \frac{\mathrm{Im}[U_{ik}] G_{kj}^{(R)}}{\sqrt{W_{kj}^{(R)}/v}} + \frac{\mathrm{Re}[U_{ik}] G_{kj}^{(I)}}{\sqrt{W_{kj}^{(I)}/v}} \right), \qquad (8)$$

where $U_{ik}$ denotes the $(i,k)$th entry of $U$. Due to the linear combination with distinct random scale variables, Eq. (8) no longer guarantees the $t$ distribution, in sharp contrast to the unitary invariance of the Wishart Kossakowski matrix.

However, our modelling guarantees universality in tail behaviours even though the entry-wise distribution is no longer exactly the $t$ distribution. Because the Gaussian numerators $G_{ij}^{(R,I)}$ in Eq. (7) have light tails, the heavy tail of a random variable $T_v$ of the $v$-parameter Student's $t$ distribution is dominated mainly by unusual small values in the chi-squared denominators $W_{ij}^{(R,I)}$. Therefore, the asymptotic tail follows a power law, $\mathbb{P}(|T_v| > x) \sim x^{-v}$, that is a regularly varying function[2], where $\mathbb{P}$ denotes the complementary cumulative distribution function (CCDF). Owing to the principle of the single big jump[29], the unitarily transformed variables $X_{ij}^{(U)}$ in Eq. (8) maintain $\mathbb{P}(|\mathrm{Re}[X_{ij}^{(U)}]| > x) \sim x^{-v}$ and $\mathbb{P}(|\mathrm{Im}[X_{ij}^{(U)}]| > x) \sim x^{-v}$ due to the negligible joint probability $\mathbb{P}(|\mathrm{Re}[X_{ij}]| > x \cap |\mathrm{Im}[X_{ij}]| > x) \sim o(x^{-v})$. This observation demonstrates that our results based on the Student's $t$ distribution correspond to universal tail behaviours for $\mathbb{P}(|T_v| > x) \sim x^{-v}$ with the asymptotic PDF $p(|x|) \sim |x|^{-(v+1)}$, regardless of the basis $\{S_k\}$.

**Liouville superoperator.** We employ column-vectorized notation to represent $N \times N$ matrix operators—including mixed states—as $N^2 \times 1$ column-vector superkets in the Liouville space $\mathcal{L}_N$. By introducing bra-flipper operators, defining an inner product on $\mathcal{L}_N$, and deriving product identities[31], we obtain the Liouville superoperator $L$, as follows:

$$L = -\frac{i}{\hbar}\{[I_N \otimes H] - [H^T \otimes I_N]\}$$
$$+ \left(\frac{\alpha}{\hbar}\right)^2 \sum_{k,l} \gamma_{kl}\left(S_k^* \otimes S_l - \frac{1}{2}(I_N \otimes S_k^\dagger S_l + S_l^T S_k^* \otimes I_N)\right), \tag{9}$$

where $I_N$ denotes the $N$-dimensional identity matrix.

**System randomness.** To characterize the randomness of a system, we utilize the Hamiltonian matrices sampled from the GUE. The matrices are obtained by $H = (X + X^\dagger)/(2N)^{1/2}$, where $X$ is sampled from a GinUE using Eq. (6). The obtained Hamiltonian satisfies the trace normalization $\langle \mathrm{Tr}[H^2] \rangle = 1/N$, where $\langle \ldots \rangle$ denotes an ensemble average.

**Perturbative sensitivity of eigenvalues.** Consider a perturbed Hamiltonian, $H = H_0 + V$, which induces the perturbation of the Liouville superoperator $\Delta L$ according to Eq. (9). When the original superoperator $L_0$ has the biorthonormal pair $\{|\rho_m^R\rangle, |\rho_m^L\rangle\}$ associated with the eigenvalue $\lambda_m$, the first- and second-order perturbation lead to the following equations:

$$\begin{aligned}\text{1st:} &\quad \Delta L |\rho_m^R\rangle + L_0|\rho_m^{R(1)}\rangle = \lambda_m |\rho_m^{R(1)}\rangle + \lambda_m^{(1)}|\rho_m^R\rangle,\\ \text{2nd:} &\quad \Delta L |\rho_m^{R(1)}\rangle + L_0 |\rho_m^{R(2)}\rangle = \lambda_m |\rho_m^{R(2)}\rangle + \lambda_m^{(1)}|\rho_m^{R(1)}\rangle + \lambda_m^{(2)}|\rho_m^R\rangle,\end{aligned} \tag{10}$$

where $|\rho_m^{R(1)}\rangle$ and $|\rho_m^{R(2)}\rangle$ denote the first- and second-order corrections to the right eigen-superket, respectively, and $\lambda_m^{(1)}$ and $\lambda_m^{(2)}$ denote the first- and second-order corrections to the eigenvalue, respectively. By applying the eigen-superbra $\langle \rho_m^L|$ to Eq. (10), $\delta\lambda_m$ in Eq. (4) is obtained.

**Haar random pure states.** To generally evaluate quantum dynamics, we employ Haar random pure states as initial states, which correspond to uniformly sampled pure states according to the Haar measure[44]. Independent states are generated by selecting the first column of independent random Haar matrices $U_\mathrm{Haar}$—the random Haar unitary evolution of the ground state, $|\psi_\mathrm{Haar}\rangle = U_\mathrm{Haar}|0\rangle$—of which the associated density operator is $\rho_\mathrm{Haar} = |\psi_\mathrm{Haar}\rangle\langle\psi_\mathrm{Haar}|$.

**Coherence time.** To estimate the coherence time $T_2$ from the time evolution $C_E(t)$, we utilize the fitting of the relative entropy to a single exponential model, as $C_E(t) = A\exp(-t/T_2) + C_0$, using nonlinear least-squares regression.

## Data availability

The data used in this study are available from the corresponding authors upon request and can also be accessed by running the code provided as Supplementary Code S1.

## Code availability

The code used in this study is based on 'Q-ROS', developed by the authors, and is provided as Supplementary Code S1 to reproduce all data presented in this paper.

## Acknowledgements

We acknowledge financial support from the National Research Foundation of Korea (NRF) through the Basic Research Laboratory (No. RS-2024-00397664), Innovation Research Center (No. RS-2024-00413957), Young Researcher Programs (No. RS-2025-00552989), and Midcareer Researcher Program (No. RS-2023-00274348), all funded by the Korean government (MSIT). This work was supported by Creative-Pioneering Researchers Program and the BK21 FOUR program of the Education and Research Program for Future ICT Pioneers in 2024, through Seoul National University. We also acknowledge administrative support from SOFT foundry institute.

## Author Contributions

All authors contributed equally to conceiving the idea, developing the code, discussing the results, and preparing the final manuscript.

## Competing Interests

The authors have no conflicts of interest to declare.

## Additional information

**Correspondence and requests for materials** should be addressed to S.Y., X.P., or N.P.

**Figure Legends**

**Fig. 1. Heavy-tailed system-environment interactions. a,** $\nu$-dependent phases of the Student's $t$ distributions determined by their tail thicknesses. **b,c,** The entire PDFs of Re[$X_{ij}$] for the $t$ distributions with different $\nu$'s and the GinUE (**b**) and their enlarged views near the tails (**c**). Dots and dashed lines in (**b,c**) denote numerical and theoretical estimations of the PDFs. The PDFs for 64 realizations are overlaid in (**b,c**). $N = 50$ for all cases.

**Fig. 2. Spectral and state properties of purely dissipative systems.** Eigenvalue distributions for GinUE (**a**), and for the ensembles generated using the Student's $t$ distributions with $\nu = 4$ (**b**), $\nu = 3$ (**c**), $\nu = 2$ (**d**), and $\nu = 1$ (**e,f**). Panels (**c-e**) employ the same plot range as panels (**a,b**), focusing on values near $\lambda = 0$; the full range plots for (**c,d**) appear in Supplementary Note S1. The colours of markers represent the $K_m$ values. For each ensemble, 64 realizations of $N = 50$ systems are examined, resulting in 160,000 eigenvalues overlaid as points on the complex plane. Blue dashed lines highlight the GinUE upper bound of eigenstate $Q$ factor by extending from the origin. We set $\alpha = 1$ and $\hbar = 1$.

**Fig. 3. Tail-dependent relaxation and sensitivity parameters of purely dissipative systems. a,b,** Relaxation parameters: spectral gaps $\Delta\lambda$ (**a**) and CCDFs of Re[$\lambda$] (**b**). **c,d,** Sensitivity parameters: the averaged Petermann factors $\langle K_m \rangle$ (**c**) and PDFs of nearest-neighbour eigenvalue spacings $\delta\lambda_{NN}$ (**d**). In (**a-d**), $\nu$-dependent heavy-tailed systems are compared with GinUE realizations. In (**a,c**), each marker represents one ensemble realization, and error bars show the median (black dashed lines), and the first and third quartiles. All the other parameters are the same as those in Fig. 2.

**Fig. 4. Coherence time and sensitivity metrics in random open systems. a,** Coherence time $T_2$. **b,c,** Sensitivity metrics: coherence time perturbation $\Delta T_2$ (**b**) and time-averaged coherence perturbation $\Delta C_E$ (**c**). Each marker represents one ensemble realization characterized by a tail-dependent $K$, a GUE Hamiltonian $H_0$ (**a**) or the pair of $H_0$ and $H_0 + V$ (**b,c**), and a Haar random pure initial state. Error bars show the median (black dashed lines), and the first and third quartiles for 1,280 realizations obtained from 64 open quantum systems and 20 initial states. All the other parameters are the same as those in Fig. 2.

169 (1933).

# Supplementary Information for "Heavy-tailed open quantum systems reveal long-lived and ultrasensitive coherence"

Sunkyu Yu[1†], Xianji Piao[2§], and Namkyoo Park[3*]

[1]Intelligent Wave Systems Laboratory, Department of Electrical and Computer Engineering, Seoul National University, Seoul 08826, Korea

[2]Wave Engineering Laboratory, School of Electrical and Computer Engineering, University of Seoul, Seoul 02504, Korea

[3]Photonic Systems Laboratory, Department of Electrical and Computer Engineering, Seoul National University, Seoul 08826, Korea

E-mail address for correspondence: [†]sunkyu.yu@snu.ac.kr, [§]piao@uos.ac.kr, [*]nkpark@snu.ac.kr

**Supplementary Note S1. MHT regime and full-range spectra in HT and EHT regimes**

**Supplementary Note S2. Spectral and state properties with GUE Hamiltonians**

**Supplementary Note S3. Re[$\lambda$] distributions**

**Supplementary Note S4. High-$Q$ states**

**Supplementary Note S5. Tail dependence with GUE Hamiltonians**

**Supplementary Note S6. Time evolutions of coherence and purity**

**Supplementary Note S1. MHT regime and full-range spectra in HT and EHT regimes**

In this note, we analyse the spectral and state properties of purely dissipative systems in the MHT regime and in the HT and EHT regimes over the entire ranges of the eigenvalue distributions at $v$ = 2 and 3. As shown in Supplementary Fig. S1a, the MHT regime at $v = 5$ exhibits almost the same eigenvalue distribution and Petermann factors as those of the GinUE realization in Fig. 2a in the main text. In the HT (Supplementary Fig. S1b) and EHT (Supplementary Fig. S1c) regimes, the maximum magnitudes of Re[$\lambda$] and Im[$\lambda$] increase substantially, leading to the emergence of much more dissipative (Re[$\lambda$] << 1) and oscillatory (large |Im[$\lambda$]|) states under CLT violation. In particular, we note that highly dissipative states correspond to outliers originating from heavy-tailed distributions, which will be discussed in Supplementary Note S3.

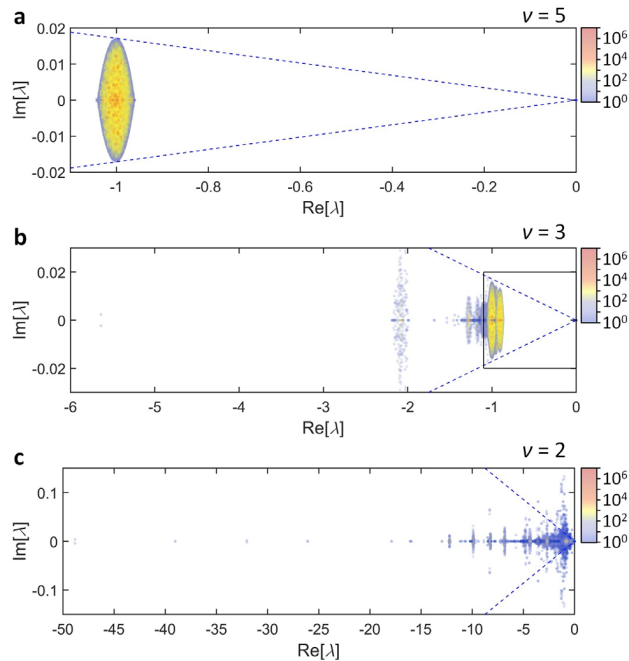

**Supplementary Figure S1. Spectral and state properties of purely dissipative systems for the MHT, HT, and EHT regimes.** Eigenvalue distributions for the ensembles generated using the $t$ distributions with $v = 5$ (**a**), $v = 3$ (**b**), and $v = 2$ (**c**). In (**b**,**c**), the entire ranges of eigenvalue distributions are plotted compared to Fig. 1c,d in the main text. The colours of markers represent the $K_m$ values. All the other parameters are the same as those in Fig. 2 in the main text.

**Supplementary Note S2. Spectral and state properties with GUE Hamiltonians**

The results for purely dissipative systems ($H = O$) described in Fig. 2 in the main text and in Supplementary Note S1 provide a clear and isolated characterization of dissipation processes in open quantum systems. On the other hand, it is necessary to investigate how such dissipation interacts with random systems that provide generalized unitary evolutions to better reflect practical configurations. Therefore, in this note, we extend the results in Fig. 2 in the main text to nonzero and random Hamiltonian $H$. As a general system under the CLT, we employ random system Hamiltonians $H$ sampled from the GUE for calculating Eq. (1) in the main text (see Methods in the main text for the realization of $H$).

Supplementary Figure S2 shows the corresponding eigenvalue and Petermann-factor distributions. With the GinUE realizations, the eigenvalue distribution undergoes a geometric transition from a "lemon-like" to an elliptic distribution boundary of which the major and minor axes are determined by the overall interaction strength[1] (Supplementary Fig. S2a). Similarly, the distributions in the HT regime with $v = 3$ and 4 (Supplementary Fig. S2b,c) closely resemble those in the purely dissipative systems shown in Fig. 2b,c in the main text, except for an overall elliptical distortion of each distribution boundary.

The most noticeable changes introduced by GUE Hamiltonians appear in the EHT regime with $v = 1$ and 2 (Supplementary Fig. S2d,e). As shown, the GUE-sampled $H$ produces a broadening of Im[$\lambda$] due to unitary evolution, leading to more oscillatory states. Remarkably, the strength of this spectral broadening substantially increases for smaller Re[$\lambda$]—lower-dissipation states. According to the perturbation theory described in Eq. (4) in the main text, the observed trend can be interpreted as the enhanced perturbative sensitivity near $\lambda = 0$. This enhanced sensitivity originates from the high density of eigenvalues in the region of smaller Re[$\lambda$], as demonstrated in Supplementary Note S3, and the consequent strong higher-order perturbations.

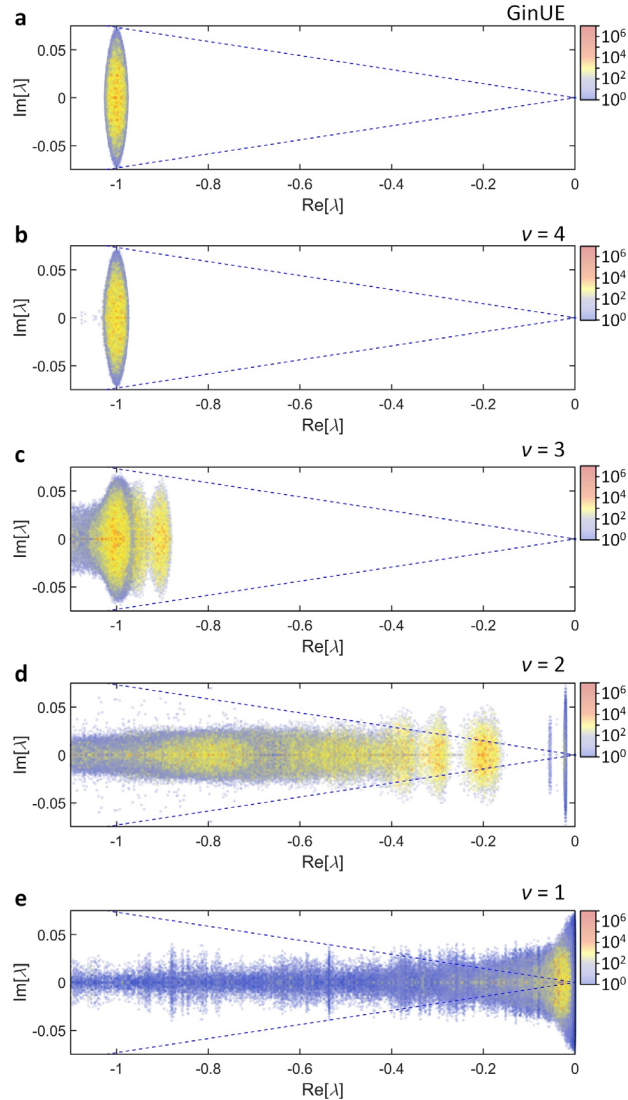

**Supplementary Figure S2. Spectral and state properties of Gaussian random open systems.** Eigenvalue distributions for GinUE (**a**), and for the ensembles generated using the Student's $t$ distributions with $v = 4$ (**b**), $v = 3$ (**c**), $v = 2$ (**d**), and $v = 1$ (**e**). The colours of markers represent the $K_m$ values. Blue dashed lines highlight the GinUE upper bound of eigenstate $Q$ factor by extending from the origin. Panels (**c-e**) employ the same plot range as panels (**a,b**), focusing on values near $\lambda = 0$; the full range plots appear in Supplementary Fig. S3. All the other parameters are the same as those in Fig. 2 in the main text.

For completeness, we also examine the spectral and state properties of Gaussian random open systems in the MHT regime, and in the HT and EHT regimes over the entire ranges of the eigenvalue distributions at $v = 1$, 2, and 3. Supplementary Figure S3a shows that the MHT regime at $v = 5$ again exhibits results nearly identical to those of the GinUE realization in Supplementary Fig. S2a, demonstrating the critical role of the CLT in understanding this heavy-tailed regime.

Notably, in the HT (Supplementary Fig. S3b with $v = 3$) and EHT (Supplementary Fig. S3c,d with $v = 2$ and 1) regimes, the states characterized by very large $|\text{Re}[\lambda]|$, which correspond to high-dissipation outliers, show almost identical features to those of purely dissipative systems described in Supplementary Fig. S1b,c and Fig. 2f in the main text. This result originates from their highly dissipative features, substantially suppressing the effect of temporal unitary evolution induced by the GUE Hamiltonian $H$.

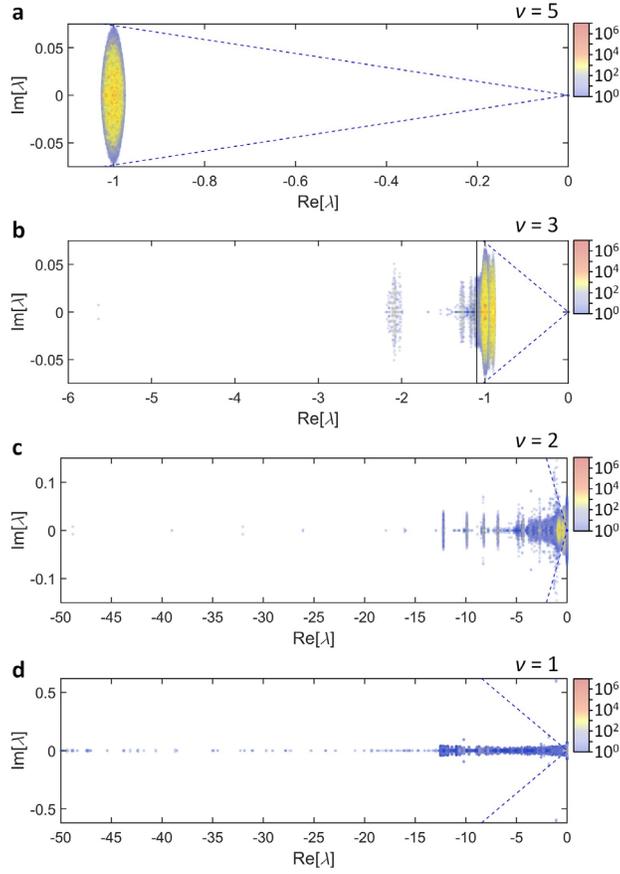

**Supplementary Figure S3. Spectral and state properties of Gaussian random open systems for the MHT, HT, and EHT regimes.** Eigenvalue distributions for the ensembles generated using the *t* distributions with $v = 5$ (**a**), $v = 3$ (**b**), $v = 2$ (**c**), and $v = 1$ (**d**). In (**b-d**), the entire ranges of eigenvalue distributions are plotted compared to Supplementary Fig. S2c-e in the main text. The colours of markers represent the $K_m$ values. All the other parameters are the same as those in Fig. 2 in the main text.

**Supplementary Note S3. Re[$\lambda$] distributions**

To investigate the origin of dissipation in random open quantum systems, we evaluate the PDFs of Re[$\lambda$] (Supplementary Fig. S4a-f) and the corresponding CCDFs (Supplementary Fig. S4g-l) for GinUE realizations and for heavy-tailed ensembles with varying $v$. In Supplementary Fig. S4, we compare purely dissipative systems (red lines) with systems governed by GUE Hamiltonians (blue lines). Consistent with Fig. 2 in the main text, the GinUE cases and the systems with $v = 5$ and 4 yield very similar results, reflecting the lemon-like shape in purely dissipative systems (Fig. 4a,b in the main text) and the elliptically distorted shape (Supplementary Fig. S2a,b) in GUE systems. In the HT regime with $v = 3$, the PDF becomes noticeably asymmetric about Re[$\lambda$] = –1, also in line with Fig. 2c in the main text.

    We note that the PDFs and CCDFs in the EHT regime with $v = 1$ and 2 are in sharp contrast to those in the HT, MHT, and GinUE cases, requiring logarithmic scales to capture their heavy-tailed nature. Although both EHT cases exhibit gap narrowing—as evidenced by large PDF values near Re[$\lambda$] = 0—the regime with $v = 1$ shows complete suppression of the localized PDF distribution near Re[$\lambda$] = –1 (Supplementary Fig. S4f), which is the signature of GinUE realizations. We also note that the alteration driven by GUE Hamiltonians is more pronounced in the $v = 1$ case (blue line with respect to red line in Supplementary Fig. S4f), which is in accordance with stronger perturbation near Re[$\lambda$] = 0 due to the higher eigenvalue density. However, the 20:80 Pareto principle, which separates high-dissipation outliers (Re[$\lambda$] < –1) and a low-dissipation bulk (Re[$\lambda$] $\geq$ –1), is valid for both purely dissipative and GUE cases, as evidenced by the nearly identical CCDF values near Re[$\lambda$] = –1 in Supplementary Fig. S4l.

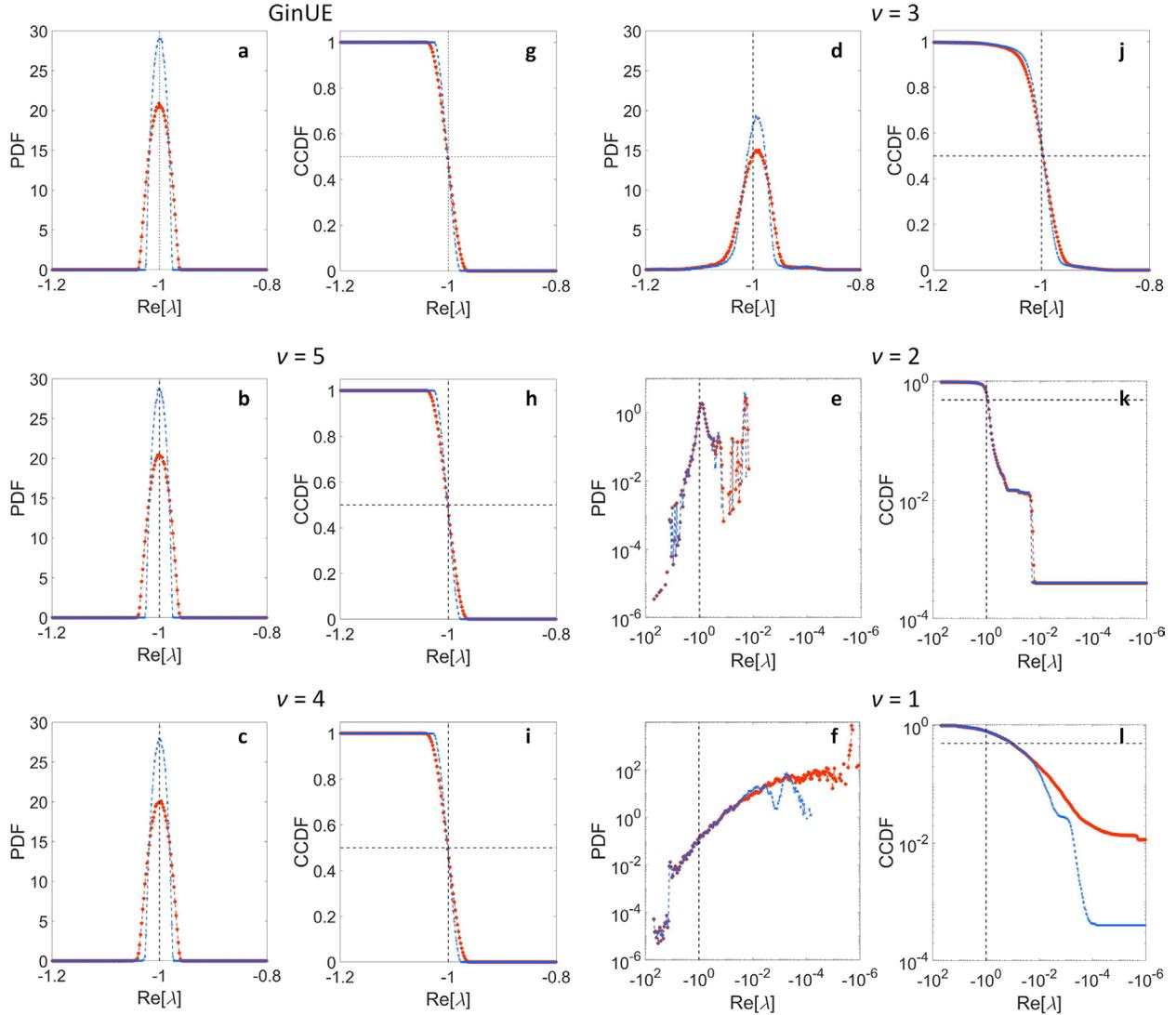

**Supplementary Figure S4. PDFs and CCDFs of Re[$\lambda$]. a-f,** PDFs and **g-l,** CCDFs for GinUE realizations (**a,g**) and for ensembles of $v = 5$ (**b,h**), $v = 4$ (**c,i**), $v = 3$ (**d,j**), $v = 2$ (**e,k**), and $v = 1$ (**f,l**). Red and blue lines denote the ensembles of purely dissipative systems ($H = O$) and of GUE Hamiltonians, respectively. The vertical black dashed lines indicate Re[$\lambda$] = –1. The horizontal black dashed lines indicate the CCDF value of 0.5. All the other parameters are the same as those in Fig. 2 in the main text.

**Supplementary Note S4. High-$Q$ states**

Supplementary Figure S5 shows the averaged $Q$ factors across the eigenstates in random open quantum systems characterized by the Student's $t$ distributions with different tail thicknesses and by the GinUE distributions. Notably, overall dissipation rises rapidly as the tails thicken (Supplementary Fig. S5a), driven by the appearance of exceptionally large magnitudes of matrix entries in the heavy-tailed Kossakowski matrix $K$ and the consequent high-dissipation outliers. This behaviour is consistent with the overall $\lambda$ distributions described in Supplementary Notes S1-S3. However, as shown in Supplementary Figs. S5b and S5c, open quantum systems with orders-of-magnitude higher $Q$ factors are obtained in the regimes of CLT violation, even exceeding the GinUE $Q$-factor bounds for the top 1% of eigenstates. Particularly in the EHT regime at $v = 1$, some systems exhibit higher $Q$-factor averages among the top 10% eigenstates.

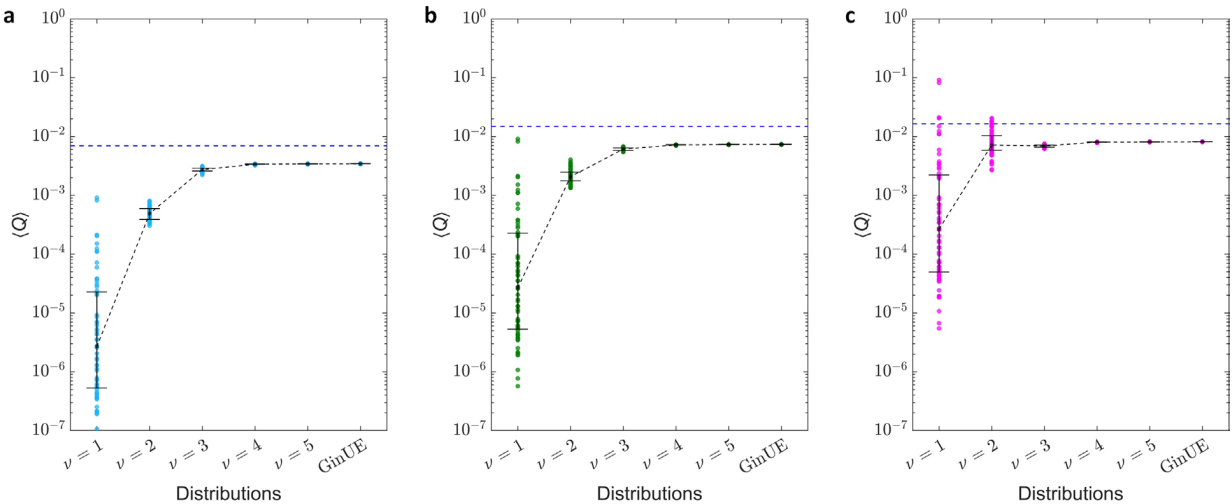

**Supplementary Figure S5. Tail-dependent $Q$ factors of eigenstates. a-c,** $v$-dependent averaged $Q$ factors of the entire eigenstates (**a**), the top 10% eigenstates (**b**), and the top 1% eigenstates (**c**), compared with GinUE realizations. Each marker represents one ensemble realization. Error bars show the median (black dashed lines), and the first and third quartiles. All the other parameters are the same as those in Fig. 2 in the main text.

**Supplementary Note S5. Tail dependence with GUE Hamiltonians**

For completeness, we analyse the tail dependence of spectral gaps, CCDFs of Re[$\lambda$], Petermann factors, and eigenvalue spacing using random system Hamiltonians $H$ sampled from the GUE in Supplementary Fig. S6. Although the unitary evolutions induced by $H$ attenuate the distinctive features of heavy-tailed open quantum systems, the general trends—decreasing spectral gaps (Supplementary Fig. S6a), 20:80 Pareto principle (Supplementary Fig. S6b), Petermann factors (Supplementary Fig. S6c), and denser eigenvalue distributions (Supplementary Fig. S6d) under CLT violation—are well preserved, which elucidate quantum dynamics described in Fig. 4 in the main text.

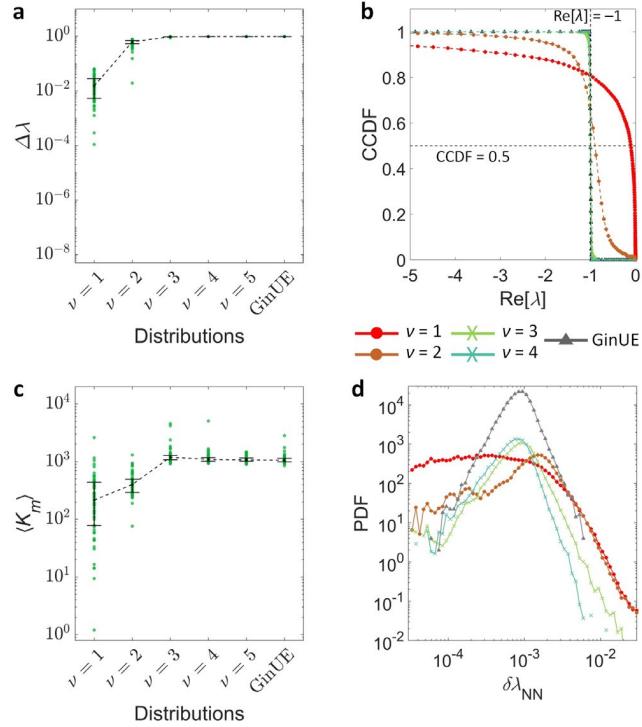

**Supplementary Figure S6. Tail-dependent relaxation and sensitivity parameters of Gaussian random open systems. a,b,** Relaxation parameters: spectral gaps $\Delta\lambda$ (**a**) and CCDFs of Re[$\lambda$] (**b**). **c,d,** Sensitivity parameters: the averaged Petermann factors $\langle K_m \rangle$ (**c**) and PDFs of nearest-neighbour eigenvalue spacings $\delta\lambda_{NN}$ (**d**). In (**a-d**), $\nu$-dependent heavy-tailed systems are compared with GinUE realizations. In (**a,c**), each marker represents one ensemble realization, and error bars show the median (black dashed lines), and the first and third quartiles. All the other parameters are the same as those in Fig. 2 in the main text.

**Supplementary Note S6. Time evolutions of coherence and purity**

In Supplementary Fig. S7, we calculate the dynamics of the coherence $C_E$ and purity $S$ in random open quantum systems with different tail thicknesses. Consistent with Figs. 2 and 3 in the main text and with the results in Supplementary Note S5, the CLT-preserved HT regimes ($v \geq 3$), which maintain the spectral gaps of the GinUE realizations and the CCDFs of Re[$\lambda$], yield quantum dynamics nearly identical to those of the GinUE realizations (Supplementary Fig. S7c,g versus S7d,h). In contrast, CLT violation leads to the emergence of long-lived quantum states (Supplementary Figs. S7a,b and S7e,f), originating from gapless spectra ($\Delta\lambda \to 0$) and 20:80 Pareto-type CCDFs.

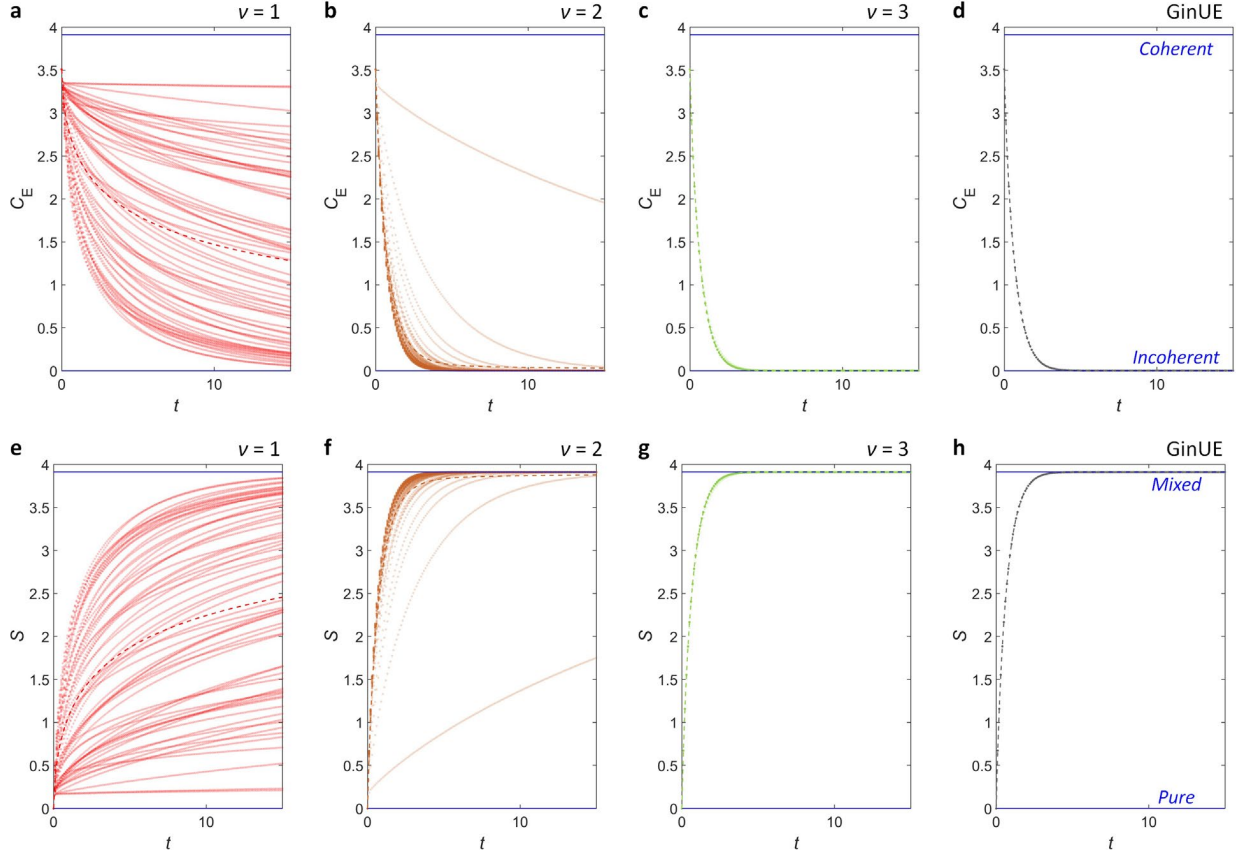

**Supplementary Figure S7. Tail-dependent dynamics in Gaussian random open systems. a-d,** Relative entropy $C_E$ for the state coherence, and **e-h,** von Neumann entropy $S$ for the state purity, which are obtained from the *t*-distribution realizations of $v = 1$ (**a,e**), $v = 2$ (**b,f**), and $v = 3$ (**c,g**), as well as the GinUE realizations (**d,h**). Each point denotes the average of the metrics for 20 initial Haar random pure states. For each ensemble, 64 realizations are plotted. Coloured dashed lines denote the ensemble-averaged dynamics. Blue solid lines indicate the accessible ranges of $C_E$ and $S$. All the other parameters are the same as those in Fig. 2 in the main text.

**Supplementary References**